# Left-censored recurrent event analysis in epidemiological studies: a proposal when the number of previous episodes is unknown

Gilma Hernández-Herra[1,2], David Moriña[3], Albert Navarro[4,5]

1. Instituto de Investigaciones Médicas, Facultad de Medicina, Universidad de Antioquia. Medellín, Colombia.

2. PhD program in Methodology of Biomedical Research and Public Health. Autnomous University of Barcelona, Cerdanyola del Vallès, Spain.

3. Department of Econometrics, Statistics and Applied Economics, Riskcenter-IREA, University of Barcelona.

4. Research group on Psychosocial Risks, Organization of Work and Health (POWAH), Autonomous University of Barcelona (UAB), Cerdanyola del Vallès, Spain

5. Biostatistics Unit, Faculty of Medicine, Autonomous University of Barcelona (UAB), Cerdanyola del Vallès, Spain

Corresponding author:

Albert Navarro. Biostatistics Unit, Faculty of Medicine, Autonomous University of Barcelona (UAB). Avda. Can Domènech S/N, 08193 Cerdanyola del Vallès, Spain. Mail: albert.navarro@uab.cat

**ABSTRACT**

Left censoring can occur with relative frequency when analysing recurrent events in epidemiological studies, especially observational ones. Concretely, the inclusion of individuals that were already at risk before the effective initiation in a cohort study, may cause the unawareness of prior episodes that have already been experienced, and this will easily lead to biased and inefficient estimates. The objective of this paper is to propose a statistical method that performs successfully in these circumstances. Our proposal is based on the use of models with specific baseline hazard, imputing the number of prior episodes when unknown, with a stratified model depending on whether the individual had or had not previously been at risk, and the use of a frailty term. The performance is examined in different scenarios through a comprehensive simulation study. The proposed method achieves notable performance even when the percentage of subjects at risk before the beginning of the follow-up is very elevated, with biases that are often under 10% and coverages of around 95%, sometimes somewhat conservative. If the baseline hazard is constant, it seems to be that the "Gap Time" approach is better; if it is not constant, the "Counting Process" seems to be a better choice. Because of the lack of knowledge of the prior episodes that have been experienced by a part (or all) of subjects, the use of common baseline methods is not advised. Our proposal seems to perform acceptably in the majority of the scenarios proposed, becoming an interesting alternative in this context.

# 1 Introduction

In cohort epidemiological studies, often all or part of the individuals included have been at risk of the event of interest before the beginning of the follow-up of the study, especially in the case of observational designs. This can lead to the unawareness of the prior history of these individuals, more specifically of the time at risk, and if they have experienced the event (if corresponds, and how many times) at the beginning of the cohort. Not knowing the amount of time an individual has been at risk of the event of interest will be a problem, when the baseline hazard of having the event depends on time. With regards to the lack of knowledge of whether the event has already happened, we can identify two situations: first, if the event can only occur once, and it has already been produced, the result for this individual is determined regardless of how long we follow him or her. We are before what is known as left censoring with a series of specific statistical techniques for its analysis; second, if the outcome of interest can be observed more than once on a same individual, in other words, it is a recurrent event, one or several new episodes of the event can be observed but the number of prior episodes will be unknown. In this case we will be in a situation of left censoring where the censored variable is of the discrete type, that can also define different baseline hazards.

This paper is situated in the context in which the prior history is unknown for all, or some, of the individuals included in a cohort, when the outcome of interest is a recurrent event with event dependence. Specifically, we suppose that we know the moment from where all individuals are at risk, we do now know the number of prior episodes they have experienced. This is a realistic situation in the practice of several cases; for example, it is very probable that in a work force cohort we know when a worker started to work (thus, to be at risk of having a sick leave) and, however, and especially for people with ample trajectory, that we don´t know whether or not in effect they have already had sick leaves (and in this case, how many). Another example, of this situation is a study of cohorts with an outcome of incidence of infection form Human Papilloma Virus on adult women. It would be relatively simple to know how long they have been at risk (beginning of active sexual life), however, we will not be able to know the number of infections since most of the time, when they occur, they are asymptomatic.

**2 Event dependence**

When we analyse a recurrent event, we often find a phenomenon called event dependency which implies that the baseline hazard of having an episode depends on the number of episodes that have already been experienced. The phenomenon of the event dependence has been estimated for several outcomes such as falls,[1] sickness absence,[2,3] hospitalizations in heart failure[4] or cardiovascular readmissions post percutaneous coronary intervention,[5] showing in all the aforementioned cases that the baseline hazard increases most significantly according to the prior episodes experienced.

Methods to analyse recurrent events in epidemiological studies are based fundamentally on extensions of Cox's[6–10] classical model of proportional hazards. Specifically, the methods that consider the existence of event dependence are specific baseline hazard models, also called conditional models or Prentice, Williams and Peterson (PWP).[11] Through stratification according to the number of prior episodes, these models assume that the baseline hazard of having an episode of the event is different as a function of the episodes that the individual has already had, allowing then to calculate a general effect or specific effects for each episode. Therefore, all individuals are at risk for the first strata, but only those with an episode on the previous strata would be at risk for the following strata.

PWP models can be formulated two different ways according to the specification of the risk interval used,[8] that is to say, according to how time is considered. In the first one, called "Counting Process", time is considered in the standard way of survival analysis, being referenced always at the beginning of the follow-up, so that the beginning of the $k_{th}$ episode is always posterior at the end of the $k\text{-}1_{st}$. Its hazard function is shown below:

$$\lambda_{ik}(t) = \lambda_{0k}(t)e^{X_i\beta} \quad (1)$$

where $X_i\beta$ represents the vector of covariables and the coefficients of the regression, $k$ is the $k_{th}$ episode of the event for individual $i$ and $\lambda_{0k}(t)$ is the function of the baseline hazard that depends on $k$.

In the second format called "Gap Time", time is considered always in relation to the previous episode,

thus being that the beginning of each new episode for a same individual is set at zero:

$$\lambda_{ik}(t) = \lambda_{0k}(t - t_{k-1})e^{X_i\beta} \quad (2)$$

If the phenomenon being studied should not present event dependence, the previous models could be "simplified" giving place to models as a function of common baseline hazard, also called Andersen-Gill Models,[12] that assign the same baseline hazard independent of the episodes that have already been experienced:

$$\lambda_i(t) = \lambda_0(t)e^{X_i\beta} \quad (3)$$

where $\lambda_0(t)$ is the common baseline hazard.

## 3 Individual heterogeneity

If a model is perfectly specified, so that all possible relevant covariates are accounted for, then the baseline hazard reflects the randomness of the event time, given the value of covariates. In practice, however, it is rarely possible to account for all relevant covariates.[13] Then, another aspect to be considered are the unmeasured effects produced by between-subject variability, presumably due to unobserved exposures. This phenomenon is called individual heterogeneity and in practice is analysed adding a frailty to the model, $v_i$, in other words, an individual random effect to account for this "extra" variability. Since $v_i$ is a multiplicative effect, we can imagine that it represents the cumulative effect of one or more omitted covariates.[14] The most commonly-adopted frailty term has $E[v_i]=1$ and $V[v_i]=\theta$.[15–17] In this context, the models specified in equations (1), (2) and (3) receive the names of "Conditional Frailty Model - Counting Process" (4), "Conditional Frailty Model - Gap Time" (5) and "Shared Frailty Model" (6):

$$\lambda_{ik}(t) = v_i\lambda_{0k}(t)e^{X_i\beta} \quad (4)$$
$$\lambda_{ik}(t) = v_i\lambda_{0k}(t - t_{k-1})e^{X_i\beta} \quad (5)$$
$$\lambda_i(t) = v_i\lambda_0(t)e^{X_i\beta} \quad (6)$$

## 4 The problem of being unaware of the previous history of some of the individuals

Specific baseline hazard methods, either with or without frailties, can be applied when all the required information is known, in particular, the number of prior episodes that each individual has had. In practice, however, this information is not always available, which prevents from including basic information to consider the event dependence.

See Figure 1. This figure represents two subjects; on the top part, according to the counting process formulation, and on the bottom part, according to gap time. Notice that finally the difference between both formulations is that gap time "restarts" time to risk at *t=0* every time an episode is produced. At the moment of starting our study, the first subject (id=1) had a considerable amount of time at risk of the event of interest and had had two episodes (this is, then, a left-censored observation). However, this information is not known, we only know that as of the moment he/she is effectively followed, he/she has two other episodes. The second one starts its exposure effectively on *t=0*, presents an episode on *t=5* and stops being followed on *t=7*. The data tables to the right show the data that would be analysed. Notice that both in the counting process as in the gap time, the previous history of individual 1 "disappears" and it would seem to be that he or she barely has two episodes and starts to be at risk exactly at the same instant as individual 2.

(Figure 1)

Thus, we find ourselves before two problems: first, if there is event dependence, and we don´t stratify by number of episode, we mix individuals that at a same instant have different baseline hazards (for example, when id=2 starts to be at risk at *t=0* for the first episode, id=1 is already at risk of the third); second, we also mix temporal scales. On one side there will be subjects, such as id=2, whose scale is follow-up time that at the same time corresponds with time at risk of the event, but on the other side there will be subjects, for example id=1, whose follow-up time does not correspond at all to his/her time at risk. As a consequence, what occurs is two subpopulations are mixed, whose baseline hazard is not the same, going against the assumptions of the Cox model and its extensions for recurrent phenomenon.

Because of the lack of knowledge of the previous history of some of the subjects, and because of the impossibility of using models of specific baseline hazard, the alternative usually is ignoring that history and adjusting models based on common baseline hazard (equations 3 and 6). However, these models assign the same baseline hazard to all episodes and, thus, do not consider the possible effect of the number of episodes that have already occurred, and it doesn't consider that it is comparing two individuals at the same instant, with times to risk that can be radically different. In fact, the use of models with a common baseline hazard for the analysis of recurrent phenomena with event dependence has been shown to be highly inefficient, generating high levels of bias in the estimate of parameters and coverages of confidence intervals, very much under what was expected, even if the event dependence is of small intensity.[18] All of this shows the need to explore analytical alternatives that may work acceptably in this context.

## 5 Objectives

This study has two objectives: first, describe an analysis proposal for recurrent phenomena in the presence of event dependence when there are subjects whose previous history is unknown; and second, compare the performance of our proposal with that of the model that ignores event dependence.

## 6 Proposal

Our proposal starts from the assumption that, even though the previous history of all or some of the individuals is unknown, we do know which of these were at risk prior to the beginning of the follow-up and starting when, and that is based fundamentally on three considerations: 1) impute $k$, the number of previous episodes for those subjects at risk before the beginning of the follow-up; 2) treat the subpopulation of subjects "Previously at risk" separately from those "Not previously at risk", and 3) use a frailty term to capture the impact of unobserved effects, including uncertainty related to the imputation process. Concretely, in the two formulations, "Counting process" (7) and "Gap time" (8), the ones we call "Frailty Specific Hazard Model Imputed", in its versions - Counting Process (SHFMI.CP) and Gap Time (SHFMI.GT) :

$$\lambda_{ikr}(t) = \upsilon_i \lambda_{0kr}(t) e^{X_i \beta} \quad (7)$$

$$\lambda_{ikr}(t) = \upsilon_i \lambda_{0kr}(t - t_{k-1}) e^{X_i \beta} \quad (8)$$

where $k$ will be the number of previous episodes of individual $i$ in case they are known or their imputed value in case they are not known; $r$ indicates the subpopulation the individual belongs to: "Previously at risk" or "Not previously at risk". In practice, this proposal means that we stratify by the interaction between having been at risk or not before the beginning of the follow-up, and the number of previous episodes.

Therefore, the use of the term individual random error $\upsilon_i$ intends to capture the error that will be made when imputing, as well as the effect of any variable that, having a non-nil effect, would not have been considered in the analysis. Stratifying by the number of prior episodes intends to safeguard the event dependence, and doing it as an interaction with the fact that it is an individual previously at risk, or not, separates the two subpopulations so as to not mix times, that are not comparable, on the same scale.

In order to carry out the imputation of previous history a generalised linear model (GLM) based on the Conway-Maxwell Poisson distribution (COMPoisson)[19] is fitted to the observed number of events. Imputed values are randomly sampled from the corresponding distribution with the parameters obtained in the previous step, including random noise generated from a normal distribution. In order to produce proper estimation of uncertainty, the described methodology is used within a multiple imputation framework. The results reported in Section Results correspond to the combination of m = 5 imputed data sets, according to the well-known Rubin's rules[20] and based on the following steps in a Bayesian context:

1. Fit the corresponding COMPoisson model and find the posterior mean $\hat{\beta}$ and variance $V(\hat{\beta})$ of model parameters $\beta$.
2. Draw new parameters $\beta^*$ from $N(\hat{\beta}, V(\hat{\beta}))$.
3. Compute predicted scores p using the parameters obtained in the previous step.

4. Draw imputations from the COMPoisson distribution and scores obtained in the previous step.

A detailed description of the whole imputation process can be found in Hernández et al. (2020).[21]

**7 Simulation study**

Six populations were simulated through the **R** package survsim,[22] all corresponding to previously described cohorts of workers.[18,23] The first three simulate the history of the workers depending on the occurrence of the long term sick leave (SL) with several intensities of event dependence. Since the simulations were carried out through Weibull distributions with ancillary parameter equal to 1 (that is, exponential distributions) the hazard is supposedly constant within each episode, but different between episodes (in other words, there is event dependence). For the fourth, fifth, and sixth populations, the worker cohorts are equally simulated and the outcome is SL according to the diagnosis (respiratory system, musculoskeletal system, and mental and behavioral disorders). In this case, the hazard functions are not constant within the episode. Table 1 shows the characteristics of each episode in each one of these populations. The maximum number of episodes that a subject may suffer was not fixed, although the baseline hazard was considered constant when $k \geq 3$. $X_1$, $X_2$, and $X_3$ are covariates that represent the exposure, with $X_i \sim Bernoulli(0.5)$, and $\beta_1=0.25$, $\beta_2=0.5$, and $\beta_3=0.75$ being their parameters that represent effects of different magnitudes, set independently of the episode $k$ to which the worker is exposed.

(Table 1)

Four different situations were simulated for each population, that are combinations between two possible follow-up times (2 and 5 years) and two maximum times at risk prior to the beginning of the cohort (2 and 10 years). For example, for the case of 10 years of maximum time at risk and 5 years of follow-up, dynamic populations with 15 years of history were generated and we selected the subjects who were either present at 10 years of follow-up or incorporated after that date. Follow-up time was then re-scaled, setting t=0 at 10 years for subjects already present in the population and t=0 at the beginning of the follow-up period for those incorporated later. This procedure allowed us to obtain a cohort in which some subjects had a work history of up to 10 years, which were treated as unknown, and with 5 years of effective follow-up (that observed between 10-15 years of the original simulated follow-up).

In each case, samples of sizes 250, 500 and 1000 were simulated with different proportions of subjects at risk prior to the beginning of the cohort (0.1, 0.3, 0.5 and 1).

Our proposal was compared to a model with frailty and common baseline hazard in terms of the number of previous events, but different as per subpopulation (to previous risk or not). We could call this model "Frailty Common Hazard Model with stratification by subpopulation" (CHFM.strata), in other words, a model that does not take event dependence into account, as the one expressed in equation (6), but that separates individuals according to whether they have been previously at risk, or not:

$$\lambda_{ir}(t)= \upsilon_i\lambda_{0r}(t)e^{X_i\beta} \quad (9)$$

## 8 Results

The results presented are the ones that refer to n=1000 and follow-up of 5 years, since the observed differences for n=250 and n=500, as well as for a follow-up of 2 years are not considered very relevant (see supplementary material).

With regards to the bias, Figure 2 highlights that the CHFM.strata model only obtains values under 10% in population 1 and in some case when the percentage of individuals at previous risk is 100%. Models SHFMI.CP and SHFMI.GT generally obtain biases under 10% in most situations, except for SHFMI.GT in populations 4 and 5, which is slightly above, and when the percentage of individuals at previous risk at t=0 is 100%, where SHFMI.CP seems to be more sensitive, especially in population 3. So, for the first three populations, SHFMI.GT shows equal or less bias than SHFMI.CP, whereas for 4, 5 and 6 it is more the opposite as long as there is at least 50% of the individuals that start their risk during the cohort.

(Figure 2)

The average length of 95%CI of the CHFM.strata model, except in population 1, is the largest of the three models for percentages of individuals at previous risk of up to 30%, always overcoming from there the preciseness of the SHFMI.CP model, as well as in the first three populations to SHFMI.GT, that is the most precise one in the last three populations, figure 3.

(Figure 3)

The coverage of 95%CI for the CHFM.strata model is clearly under 95% in most cases, and is lower still, the higher the event dependence; and, for the first four populations, the percentage of individuals at previous risk prior to t0. The coverages for models SHFMI.CP and SHFMI.GT are generally acceptable, even in some case, excessively conservative (over 95%). The SHFMI.CP model fails in excess for population 3 when there is 100% of the individuals at previous risk because of the high bias found, whereas SHFMI.GT obtains coverages even under 80% when in population 4 the percentage of individuals at previous risk is lower than 100%, figure 4.

(Figure 4)

Tables S1 and S2 (supplementary material) summarize the results in terms of bias and coverage using arbitrary criteria (<10% of relative bias and covergae between 92.5% and 97.5%). Considering all possible situations, CHFM.strata rarely meets the criteria, while SHFMI.CP is the model with higher level of compliance, followed by SHFMI.GT.

## 9 Final remarks

From the results obtained we can make some considerations and suggestions:

1. Not having availability of the number of previous episodes that an individual has had should not be a justification for the use of models with a common baseline hazard, that on the large majority of occasions show a higher bias and less coverage than specific baseline hazard models.
2. The first step, indispensable for a correct choice of the method to be used, should be the description of the baseline hazard form of the phenomenon being studied in each one of the episodes.
3. If the phenomenon of interest is generated from a function of constant risk, the best model to choose

should be SHFMI.GT, if there are percentages of up to 30-50% of lack of information in terms of the number of previous episodes experienced, model SHFMI.CP would also be more than acceptable.

4. For phenomenon ruled by non-constant hazard functions, model SHFMI.CP should be selected for percentages up to 50% of lack of previous information.

5. If the number of previous episodes experienced is unknown for all the individuals, although SHFMI.GT could be adequate in some cases, neither model appears to be a robust option.

6. The implementation of the proposal presented in this article is already working on standard software and ready to be used by any user that could be interested. Specifically, in **R** package `miRecSurv`.[24]

**Acknowledgements**

D. Moriña acknowledges financial support from the Ministry of Economy and Competitiveness, through the María de Maeztu Programme for Units of Excellence in R&D (MDM-2014-0445) and Fundación Santander Universidades.

**Table 1. Characteristics of the simulated populations.**

| Episode | Distribution | $\beta_0$ | Ancillary | HR |
|---|---|---|---|---|
| 1 | Weibull | 8.109 | 1 | 1 |
| 2 | Weibull | 7.927 | 1 | 1.20 |
| ≥ 3 | Weibull | 7.745 | 1 | 1.44 |
| 1 | Weibull | 8.109 | 1 | 1 |
| 2 | Weibull | 7.703 | 1 | 1.50 |
| ≥ 3 | Weibull | 7.298 | 1 | 2.25 |
| 1 | Weibull | 8.109 | 1 | 1 |
| 2 | Weibull | 7.193 | 1 | 2.50 |
| ≥ 3 | Weibull | 6.276 | 1 | 6.25 |
| 1 | Log-normal | 7.195 | 1.498 | 1 |
| 2 | Log-logistic | 6.583 | 0.924 | 1.77 |
| ≥ 3 | Weibull | 6.678 | 0.923 | 2.53 |
| 1 | Log-logistic | 7.974 | 0.836 | 1 |
| 2 | Weibull | 7.109 | 0.758 | 3.81 |
| ≥ 3 | Log-normal | 5.853 | 1.989 | 7.19 |
| 1 | Log-normal | 8.924 | 1.545 | 1 |
| 2 | Log-normal | 6.650 | 2.399 | 10.13 |
| ≥ 3 | Log-normal | 6.696 | 2.246 | 11.19 |

**Figure 1. Example of the history of two individuals, according to counting process and gap time formulation.**

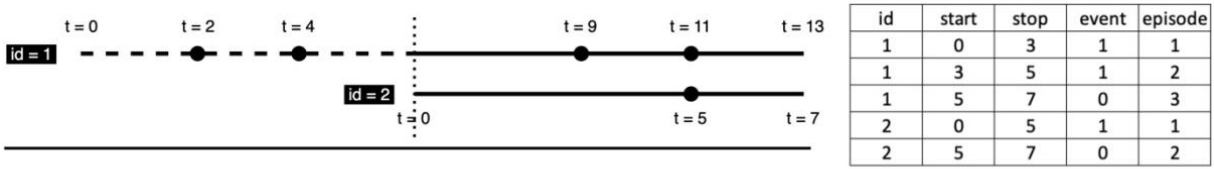

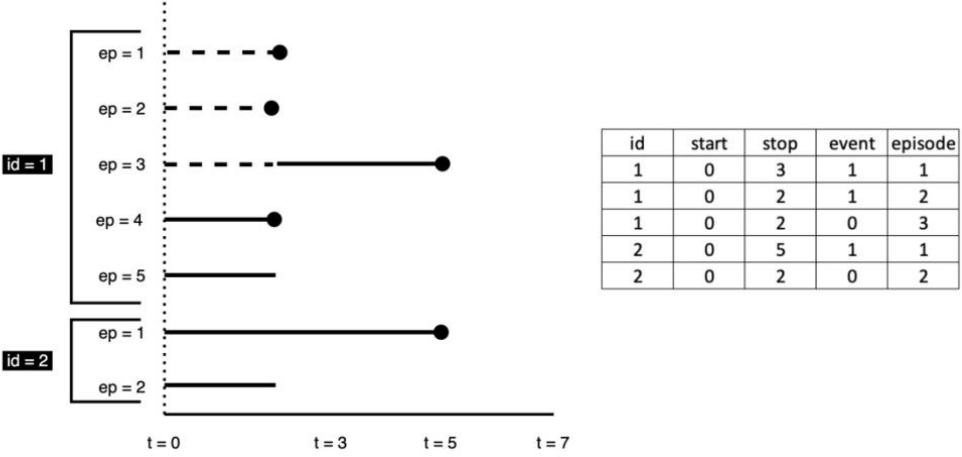

**Figure 2. Bias according to population and time of follow-up.**

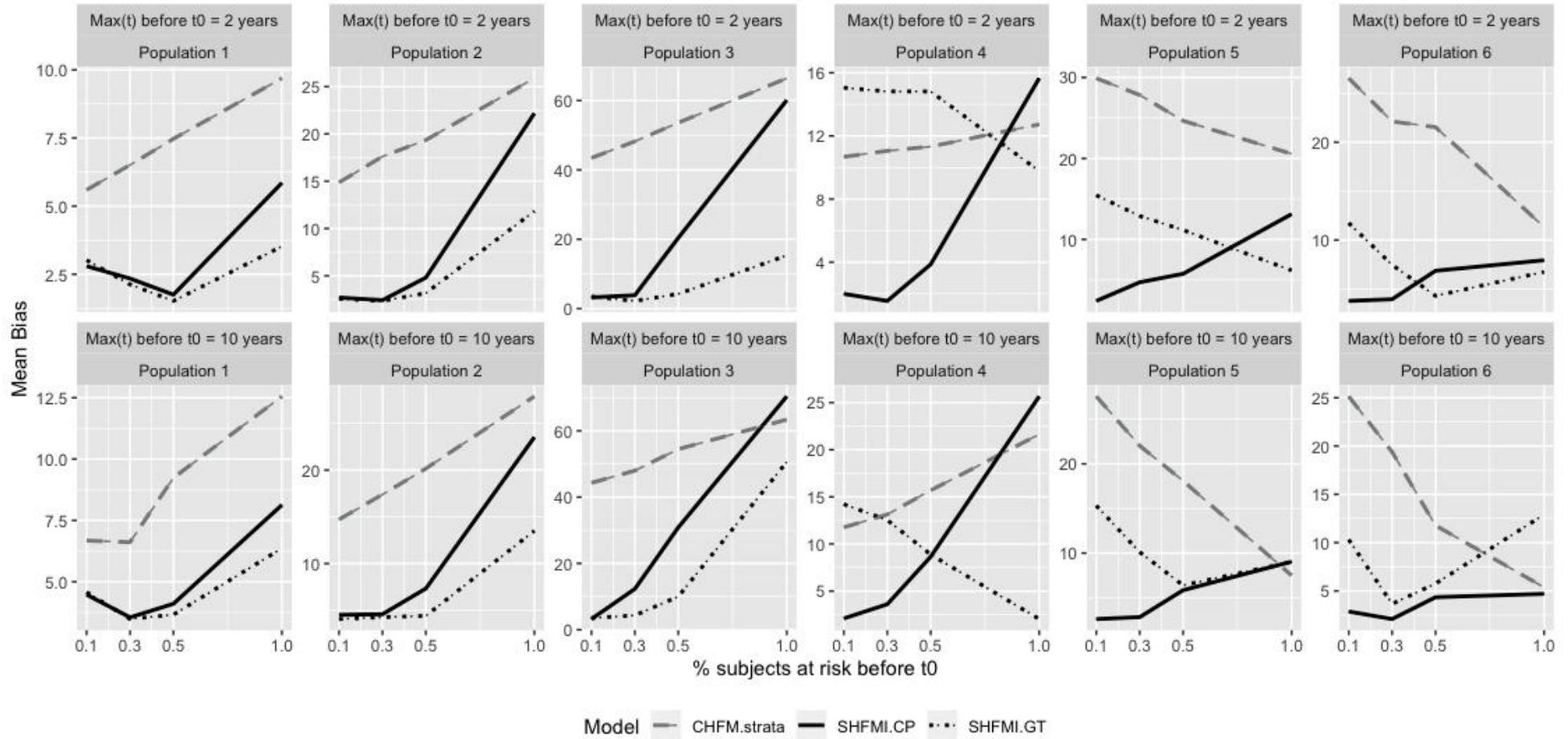

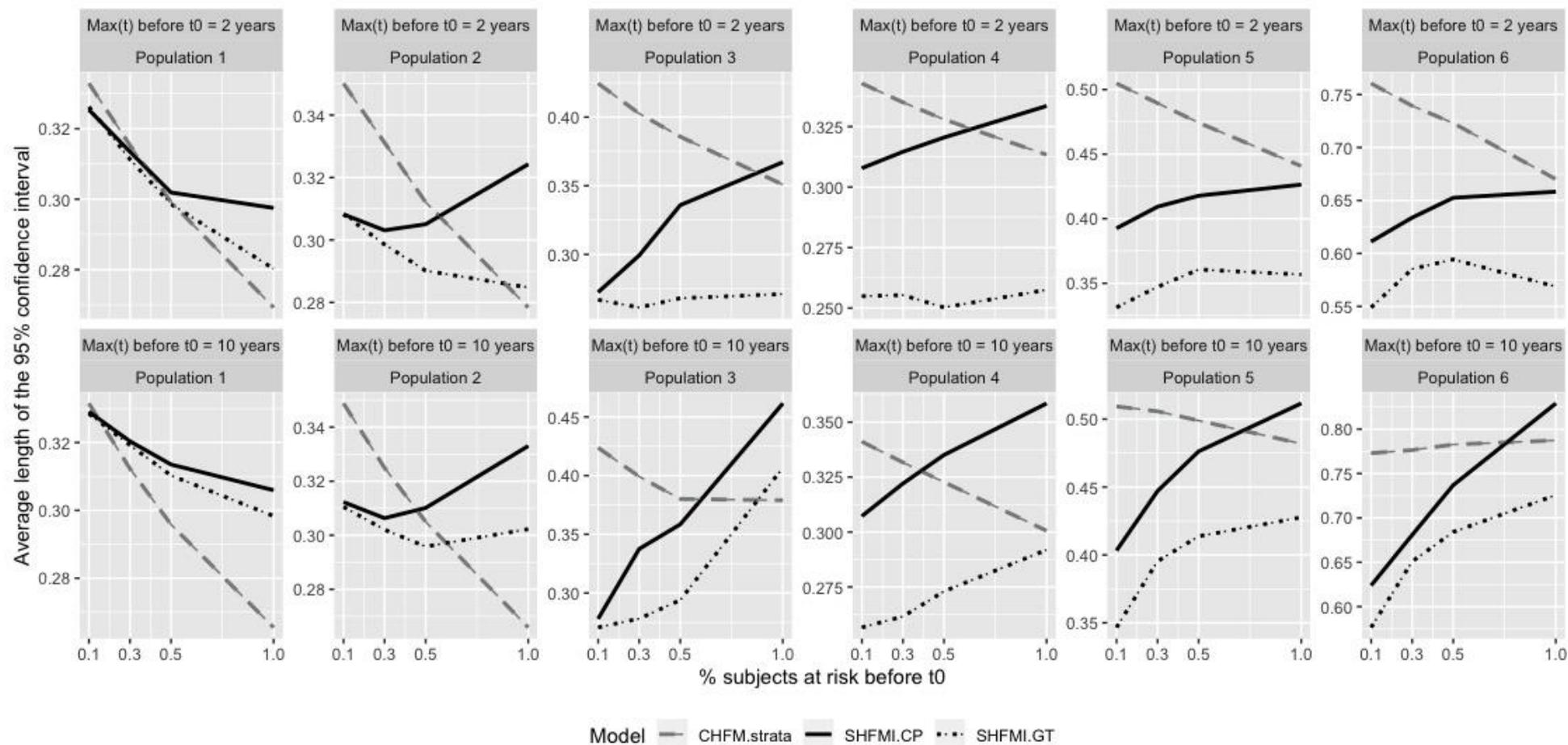

Figure 3. Average length of the 95% confidence interval according to population and time of follow-up.

**Figure 4. Coverage according to population and time of follow-up.**

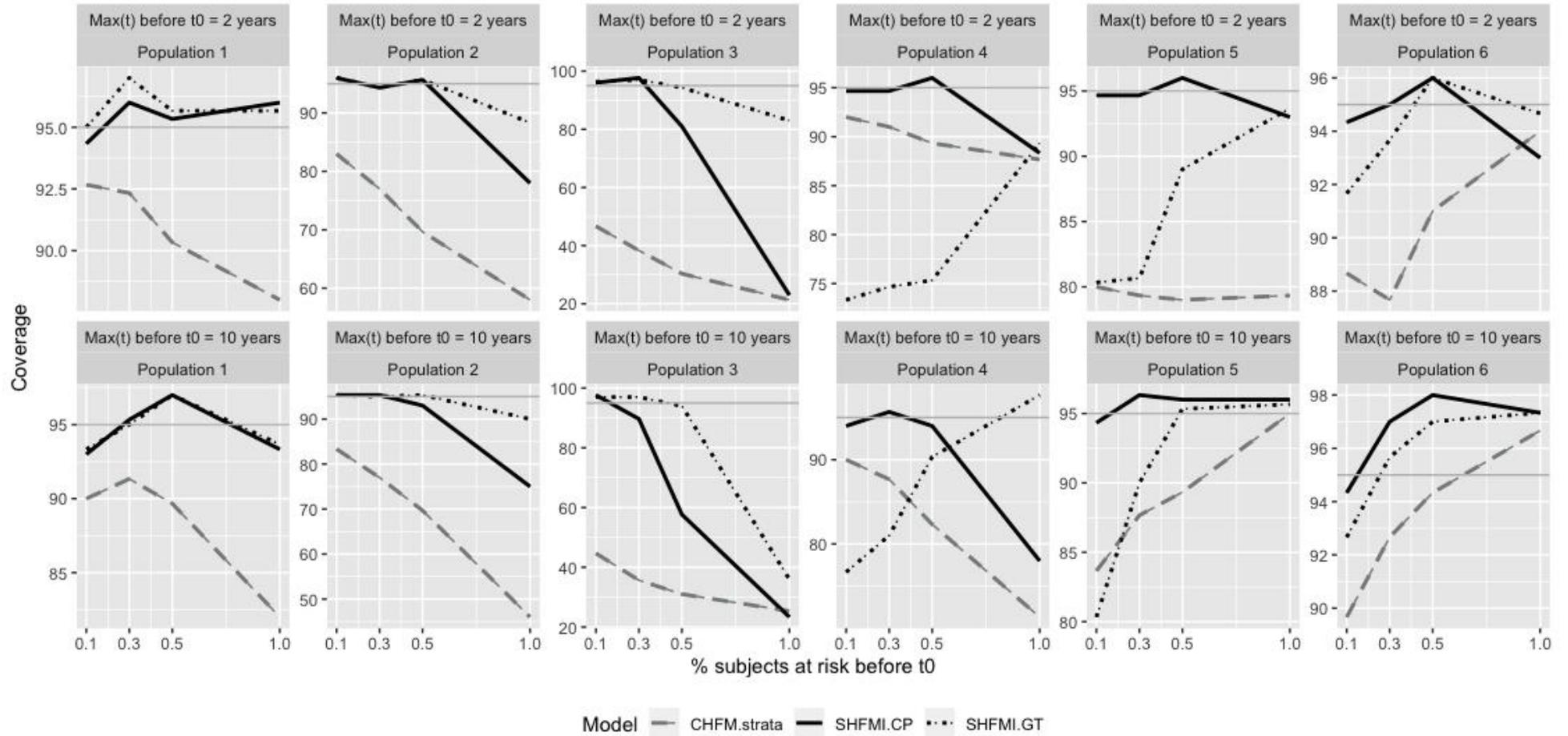

**Table S1. Relative bias < 10%.**

| | Population 1 | | | | Population 2 | | | | Population 3 | | | | Population 4 | | | | Population 5 | | | | Population 6 | | | |
|---|---|---|---|---|---|---|---|---|---|---|---|---|---|---|---|---|---|---|---|---|---|---|---|---|
| Subjects at risk before t0 | 0.1 | 0.3 | 0.5 | 1 | 0.1 | 0.3 | 0.5 | 1 | 0.1 | 0.3 | 0.5 | 1 | 0.1 | 0.3 | 0.5 | 1 | 0.1 | 0.3 | 0.5 | 1 | 0.1 | 0.3 | 0.5 | 1 |
| *Max(t) before t0 = 2* | | | | | | | | | | | | | | | | | | | | | | | | |
|   CHFM.strata | ✓ | ✓ | ✓ | ✓ | | | | | | | | | | | | | | | | | | | | |
|   SHFMI.CP | ✓ | ✓ | ✓ | ✓ | ✓ | ✓ | ✓ | | ✓ | ✓ | | | ✓ | ✓ | ✓ | | ✓ | ✓ | ✓ | | ✓ | ✓ | ✓ | ✓ |
|   SHFMI.GT | ✓ | ✓ | ✓ | ✓ | ✓ | ✓ | ✓ | | ✓ | ✓ | ✓ | | | | ✓ | | | | ✓ | | | ✓ | ✓ | ✓ |
| *Max(t) before t0 = 10* | | | | | | | | | | | | | | | | | | | | | | | | |
|   CHFM.strata | ✓ | ✓ | ✓ | | | | | | | | | | | | | | | | | | | | | |
|   SHFMI.CP | ✓ | ✓ | ✓ | ✓ | ✓ | ✓ | ✓ | | ✓ | ✓ | | | ✓ | ✓ | ✓ | | ✓ | ✓ | ✓ | ✓ | ✓ | ✓ | ✓ | ✓ |
|   SHFMI.GT | ✓ | ✓ | ✓ | ✓ | ✓ | ✓ | ✓ | | ✓ | ✓ | ✓ | | | | ✓ | ✓ | ✓ | ✓ | ✓ | | | ✓ | ✓ | |

**Table S2. Coverage between 92.5% and 97.5%.**

| | Population 1 | | | | Population 2 | | | | Population 3 | | | | Population 4 | | | | Population 5 | | | | Population 6 | | | |
|---|---|---|---|---|---|---|---|---|---|---|---|---|---|---|---|---|---|---|---|---|---|---|---|---|
| Subjects at risk before t0 | 0.1 | 0.3 | 0.5 | 1 | 0.1 | 0.3 | 0.5 | 1 | 0.1 | 0.3 | 0.5 | 1 | 0.1 | 0.3 | 0.5 | 1 | 0.1 | 0.3 | 0.5 | 1 | 0.1 | 0.3 | 0.5 | 1 |
| *Max(t) before t0 = 2* | | | | | | | | | | | | | | | | | | | | | | | | |
|   CHFM.strata | ✓ | ✓ | | | | | | | | | | | | | | | | | | | | | | ✓ |
|   SHFMI.CP | ✓ | ✓ | ✓ | ✓ | ✓ | ✓ | ✓ | | ✓ | ✓ | | | ✓ | ✓ | ✓ | | ✓ | ✓ | ✓ | ✓ | ✓ | ✓ | ✓ | ✓ |
|   SHFMI.GT | ✓ | ✓ | ✓ | ✓ | ✓ | ✓ | ✓ | | ✓ | ✓ | ✓ | | | | | | | | ✓ | | | ✓ | ✓ | ✓ |
| *Max(t) before t0 = 10* | | | | | | | | | | | | | | | | | | | | | | | | |
|   CHFM.strata | | | | | | | | | | | | | | | | | | | | ✓ | | ✓ | ✓ | ✓ |
|   SHFMI.CP | ✓ | ✓ | ✓ | ✓ | ✓ | ✓ | ✓ | | ✓ | | | | ✓ | ✓ | ✓ | | ✓ | ✓ | ✓ | ✓ | ✓ | ✓ | | ✓ |
|   SHFMI.GT | ✓ | ✓ | ✓ | ✓ | ✓ | ✓ | ✓ | | ✓ | ✓ | ✓ | | | | | ✓ | | ✓ | ✓ | | ✓ | ✓ | ✓ | ✓ |